\documentclass[12pt,preprint]{aastex}

\usepackage{epsfig}

\newcommand{\be}{\begin{equation}}
\newcommand{\ee}{\end{equation}}

\newcommand{\bea}{\begin{eqnarray}}
\newcommand{\eea}{\end{eqnarray}}

\newcommand{\al}{\alpha}

\newcommand{\Gm}{\Gamma}
\newcommand{\dl}{\delta}
\newcommand{\Dl}{\Delta}
\newcommand{\eps}{\epsilon}

\newcommand{\Lm}{\Lambda}

\newcommand{\rh}{\rho}

\newcommand{\sg}{\sigma}

\newcommand{\Om}{\Omega}

\newcommand{\rarrow}{\rightarrow}
\newcommand{\Rarrow}{\Rightarrow}

\newcommand{\nn}{\nonumber}

\newcommand{\varep}{\varepsilon}

\begin{document}

\title{The dark energy as a natural property of cosmic polytropes - A tutorial}

\author{Kostas Kleidis $^1$* and Nikolaos K. Spyrou $^2$}

\affil{$^{1}$ Department of Mechanical Engineering, International Hellenic University (Serres Campus), 621.24 Serres, Greece \\ $^{2}$ Department of Astronomy, Aristotle University of Thessaloniki, 541.24 Thessaloniki, Greece}




\begin{abstract} 

Theoretical results on a conventional approach to the dark energy (DE) concept are reviewed and discussed. According to them, there is absolutely no need for a novel DE component in the Universe, provided that the associated matter-energy content is represented by a perfect fluid whose volume elements perform polytropic flows. When the thermodynamic energy of this fluid's internal motions is also considered as a source of the universal gravitational field, it compensates the DE needed to compromise spatial flatness in an accelerating Universe. The cosmological model with matter-energy content in the form of a polytropic fluid not only interprets the observations associated to the recent history of Universe expansion, but successfully confronts with all the current cosmological issues, thus arising as a viable alternative to $\Lm$CDM model.

\end{abstract}









\section{Introduction}

According to a considerable amount of observational data accumulated in the last 25 years, it became evident that a uniformly distributed energy component, the so-called DE, is present in the Universe (see, e.g., [1, 2]). First, it was the high-precision distance measurements, performed with the aid of distant Supernova Type Ia (SNe Ia) events, which revealed that, in a dust Universe (i.e., under the assumption that the constituents of the Universe matter content do not interact with each other, so as their world lines remain eternally parallel), these standard candles look fainter (i.e., they are located farther) than what was theoretically predicted [3 - 31]. To interpret this result, Perlmutter et al. [2] and Riess et al. [9], following Carroll et al. [32], admited that the long sought cosmological constant, $\Lm$, differs from zero; hence, apart from matter, the Universe contains also a uniformly distributed amount of energy [33]. The need for an energy component that does not cluster at any scale was subsequently verified by observations of galaxy clusters [34], the integrated Sachs-Wolfe effect [35], baryon acoustic oscillations (BAOs) [36, 37], weak gravitational lensing [38, 39], and the Lyman-$\al$ forest [40]. If this energy component is due to the cosmological constant, it would necessarily introduce a repulsive gravitational force [41]; hence, the unexpected dimming of the SNe Ia standard candles was accordingly attributed to a recent acceleration of the Universe expansion (see, e.g., [42, 43]).

At the same time, high precision cosmic microwave background (CMB) observations suggested that our Universe is, in fact, a spatially-flat Robertson-Walker (RW) cosmological model [45 - 56]. This means that the overall energy density, $\varep$, of the Universe matter-energy content, in units of the critical energy density, $\varep_c = \rh_c c^2$ (the equivalent to the critical rest-mass density, $\rh_c= \frac{3 H_0^2}{8 \pi G}$, where $H_0$ is the Hubble parameter at the present epoch, $G$ is Newton's gravitational constant, and $c$ is the velocity of light), must be equal to unity, $\Om = \frac{\varep}{ \varep_c} = 1$, i.e., much larger than the measured value of the mass-density parameter, $\Om_M = \frac{\rh}{\rh_c} = 0.302 \pm 0.006$, where $\rh$ is the rest-mass density [57]. Therefore, an extra amount of energy was also needed, to justify spatial flatness.

Quantum vacuum could serve as such an energy basin, attributing an effective cosmological constant to the Universe, which would justify both spatial flatness and accelerated expansion [33], [41], [58]. Unfortunately, vacuum energy is $10^{123}$ times larger than the associated measured quantity in curved spacetime [58]. Clearly, an approach other than the cosmological constant (namely, the DE) was needed to incorporate spatial flatness in an accelerating Universe; hence, (too) many models were proposed. An (only-) indicative list would involve quintessence [59], k-essence [60], and other (more exotic) scalar fields [61], tachyons [62], brane cosmology [63, 64], scalar-tensor gravity [65], $f(R)$-theory [66, 67], holographic principle [68 - 70], Chaplygin gas [71 - 74], Cardassian cosmology [75 - 77], multidimensional cosmology [78 - 81], mass-varying neutrinos [82, 83], cosmological principle deviations [84 - 87], and many other models (see, e.g., [8]), not to mention the associated cosmographic results [89 - 108]. In an effort to illuminate darkness, we point out that, long before the necessity of DE's invention, another dark component was (and still is) present in the composition of the Universe matter content, the long sought dark matter (DM).

Today, there is absolutely no doubt as regards the existence of a non-luminous mass component in the Universe. The associated observational data involve high-precicion measurements of the flattened galactic rotation curves [109, 110], weak gravitational lensing (WGL) [111], and modulation of the strong lensing effects due to massive elliptical galaxies [112]. On galactic scale, it was found that their dark haloes extend almost half the distance to the neighboring cosmic structures [113, 114], while, at even larger scales, the total-mass of galaxy clusters is proved to be tenfold as compared to their baryonic mass [115 - 117]. The same is also true at the Universe level, as it is inferred from the combination of CMB observations [53] and light-chemicals' abundances [118]. In view of all the above, it is now well established that $85 \%$ of the Universe mass content is non-luminous [119].

The precise nature of DM constituents is still unknown. There are many candidates, from ordinary stellar-size black holes, to Bose-Einstein condensates and ultralight axions [120]. Another interesting candidate are the weakly interacting massive particles (WIMPs) [121 - 123], which can be relevant to a potential detection of DM, because they annihilate through standard-model channels [124, 125]. However, regarding WIMPs, only weak-scale physics is involved, and, therefore, we argued that, practically, they do not interact with each other. Nevertheless, a few years ago, particle detectors [126, 127] and the Wilkinson Microwave Anisotropy Probe (WMAP) [128] have revealed an unexpected excess of cosmic positrons, which might be due to WIMPs collisions (see, e.g., [129 - 139]). In other words, WIMPs can be slightly collisional [140 - 144]. 

A cosmological model of self-interacting matter content could in fact unify DM and DE between them [145 - 158]. In this framework, Kleidis and Spyrou [159 - 163] admitted that the potential collisions of WIMPs maintain a tight coupling between them and their kinetic energy is re-distributed. On this assumption, the DM itself acquires fluid-like properties, and, hence, the Universe evolution is now driven by a fluid whose volume elements perform hydrodynamic flows (and not by dust). In our defense, the same assumption has been used also in modeling dark galactic haloes, significantly improving the corresponding velocity dispersion profiles [164 - 170]. If this is the case, the thermodynamic energy of the DM fluid internal motions should also be considered as a component of the Universe matter energy content that drives cosmic expansion. We cannot help but wondering, whether it could also compensate for the extra DE needed to compromise spatial flatness or not. 

This review article is organized as follows: In Section 2, we consider a spatially-flat cosmological model whose evolution is driven by a (perfect) fluid of DM, the volume elements of which perform polytropic flows [160 - 163]. Accordingly, an extra energy amount - the energy of internal motions -  arises naturally and compensates the extra DE needed to compromise spatial flatness. Such a cosmological model involves a free parameter, the associated polytropic exponent, $\Gm$. In the case where $\Gm < 1$ the cosmic pressure becomes negative and the Universe accelerates its expansion below a particular value of the cosmological redshift parameter, $z$, the so-called {\em transition redshift}, $z_{tr}$. In Section 3, we demonstrate that the polytropic DM model so assumed can confront with all the major issues of cosmological significance, since, in the constant pressure (i.e., $\Gm = 0$) limit, it fully reproduces all the predictions and the associated observational results concerning the {\em infernous} $\Lm$CDM model [160 - 162]. Finally, we conclude in Section 4.

\section{Polytropic flows in a cosmological DM fluid}

CMB has been proved a most valuable tool for reliable cosmological observations  (see, e.g., [45 - 56]). At the present epoch, data arriving from various CMB probes strongly suggest that the Universe can be described by a spatially-flat RW model, i.e., \be ds^2 = c^2 dt^2 - S^2 (t) \left ( dx^2 + dy^2 + dz^2 \right ) \: , \ee where $S(t)$ is the scale factor as a function of cosmic time, $t$. The evolution of the cosmological model given by Eq. (1) depends on the exact form and the properties of its matter-energy content. 

According to Kleidis \& Spyrou [159 - 163], in a Universe filled with interactive DM there is absolutely no need for an extra DE component. Indeed, provided that the collisions of the DM constituents are frequent enough, they can maintain a tight coupling between them so as their kinetic energy to be re-distributed. In this case, the Universe matter content acquires thermodynamic properties and the curved spacetime evolution is driven by a perfect (DM) fluid, instead of presureless dust [159]. Due to the cosmological principle, this fluid is practically homogeneous and isotropic at large scale, and, therefore, its pressure, $p$, obeys an EoS of the form $p = f (\rh)$ [160]. Now, the fundamental units of the Universe matter content are the volume elements of this (DM) fluid, i.e., closed thermodynamical systems with conserved number of particles [171]. Their motion in the interior of the cosmic fluid under consideration is determined by the conservation law \be T_{\; \; ; \nu}^{\mu \nu} = 0 \; , \ee where Greek indices refer to the four-dimensional spacetime, Latin indices refer to the three-dimensional space, the semicolon denotes covariant derivative, and $T^{\mu \nu}$ is the energy-momentum tensor of the source that drives the Universe evolution. In the particular case of a perfect fluid, $T^{\mu \nu}$ reads \be T^{\mu \nu} = (\varep + p) u^{\mu} u^{\nu} - p g^{\mu \nu} \: , \ee where $u^{\mu}$ is the four-velocity $\left ( u_{\mu}u^{\mu} = 1 \right )$, $g^{\mu \nu}$ is the Universe metric tensor, and $\varep$ is the total energy density of the fluid, which, now, is decomposed to \be \varep = \eps (\rh, T) + \rh \: U (T) \ee (see, e.g., [172], pp. 81 - 84 and 90 - 94). In Eq. (4), $T$ is the absolute temperature, $U (T)$ is the energy of this fluid's internal motions, and $\eps (\rh, T)$ represents all forms of energy besides that of internal motions. In view of Eq. (4), Eqs. (2) represent the hydrodynamic flows of volume elements in the interior of a perfect-fluid source as they are traced by an observer comoving with cosmic expansion in a maximally symmetric cosmological model (see, e.g., [173], p. 91). The evolution of such a model (see, e.g., [173] pp. 61, 62) can be determined by the Friedmann equation of the classical Friedmann-Robertson-Walker (FRW) cosmology \be H^2 = \frac{8 \pi G}{3 c^2} \varep \: , \ee where \be H = \frac{\dot{S}}{S} \ee is the Hubble parameter in terms of $S(t)$ and the dot denotes differentiation with respect to cosmic time. To solve Eq. (5), first we need to determine $\varep$, in other words $\eps$ and $U$. To do so, we can use the first law of thermodynamics in curved spacetime, \be d U + p d \left ( \frac{1}{\rh} \right ) = {\cal C} d T \ee (see, e.g., [172], p. 83), where ${\cal C}$ is the specific heat of the cosmic fluid, in connection with the zeroth component of Eq. (2), i.e., the continuity equation \be \dot{\varep} + 3 \frac{\dot{S}}{S} (\varep + p) = 0 \: . \ee Finally, we need to decide on the form of the pressure as a function of $\rh$. Accordingly, we admit that the volume elements of the Universe matter content perform polytropic flows [160 - 163].

Polytropic process is a reversible thermodynamic process in which the specific heat of a closed system evolves in a well-defined manner (see, e.g., [174], p. 2). For ${\cal C} = constant$, the system possesses only one independent state variable, the rest-mass density, and the EoS for a perfect fluid, $ p \propto \rh T $ , results in \bea p & = & p_0 \left ( \frac{\rh}{ \rh_0} \right )^{\Gm} \\ T & = & T_0 \left ( \frac{\rh}{\rh_0} \right )^{\Gm - 1} \eea (see, e.g., [160]), where $p_0$, $\rh_0$, and $T_0$ denote the present-time values of pressure, rest-mass density, and temperature, respectively, and $\Gm$ is the polytropic exponent. In such a model, Eq. (7) yields \be U = U_0 \left ( \frac{\rh}{ \rh_0} \right )^{\Gm - 1} , \ee where \be U_0 = {\cal C} T_0 + \frac{1}{\Gm - 1} \frac{p_0}{\rh_0} \ee is the present-time value of the cosmic fluid internal energy. In view of Eqs. (4) and (11), Eq. (8) is written in the form \be \Gm U_0 \left ( \dot{\rh} + 3 \frac{\dot{S}}{S} \rh \right ) + \dot{\eps} + 3 \frac{\dot{S}}{S} \eps - 3 (\Gm - 1) \rh_0 {\cal C} T_0 \frac{\dot{S}}{S} \left ( \frac{\rh}{\rh_0} \right )^{\Gm} = 0 \: . \ee Since the total number of particles in a closed system (volume element) is conserved, we furthermore have \be \dot{\rh} + 3 \frac{\dot{S}}{S} \rh = 0 \Rarrow \rh = \rh_0 \left ( \frac{S_0}{S} \right )^3 \ee and, therefore, Eq. (13) results in \be \eps = \rh_0 c^2 \left ( \frac{S_0}{S} \right )^3 - \rh_0 {\cal C} T_0 \left ( \frac{S_0}{S} \right )^{3 \Gm} . \ee By virtue of Eqs. (11) - (15), the total energy density (4) of the polytropic DM model under consideration is written in the form \be \varep = \rh_0 c^2 \left ( \frac{S_0}{S} \right )^3 + \frac{p_0}{\Gm - 1} \left ( \frac{S_0}{S} \right )^{3 \Gm} = \rh c^2 + \frac{1}{\Gm -1} \: p \ee and the Friedmann equation (5) results in \be \left ( \frac{H}{H_0} \right )^2 = \Om_M \left ( \frac{S_0}{S} \right )^3 \left [ 1 + \frac{1}{\Gm - 1} \frac{p_0}{\rh_0 c^2} \left ( \frac{S_0}{S} \right )^{3 (\Gm - 1)} \right ] . \ee Extrapolation of Eq. (17) to the present epoch, yields the corresponding value of the polytropic DM fluid pressure, i.e., \be p_0 = \rh_0 c^2 (\Gm - 1) \frac{1 - \Om_M}{\Om_M} \: . \ee In view of Eq. (18), for $\Gm < 1$, the pressure (9) is negative and so might be the quantity $\varep + 3 p$, something that would lead to $\ddot{S} > 0$ (see, e.g., [43]). In other words, for $\Gm < 1$, the polytropic DM model under consideration can accelerate its expansion. At the same time, Eq. (16) reads \be \varep = \rh_c c^2 \left [ \Om_M \left ( \frac{S_0}{S} \right )^3 + (1 - \Om_M) \left ( \frac{S_0}{S} \right )^{3 \Gm} \right ] \: , \ee the extrapolation of which to the present epoch suggests that the total energy density parameter of the polytropic DM model under consideration is exactly unity, i.e., \be \Om_0 =  \frac{\varep_0}{\varep_c} = \frac{\rh_c c^2}{\rh_c c^2} \left [ \Om_M + (1 - \Om_M) \right ] = 1 \: . \ee We see that, the polytropic DM model with $\Gm < 1$ might be an excellent (conventional) solution to the DE issue, by compromising both spatial flatness $(\Om_0 = 1)$ and accelerated expansion $( \varep + 3 p < 0)$ of the Universe in a unique theoretical framework. 

\section{Predictions and outcomes of the polytropic DM model}

In this Section, we explore the properties of a polytropic DM model with $\Gm < 1$, in association to all the major issues of cosmological significance. To do so, unless otherwise is stated, in what follows we admit that $\Om_M = 0.274$, as suggested by the {\it nine years WMAP survey} [54]. This value differs from the corresponding {\it Planck} result, $\Om_M = 0.308$ [55, 56], and/or the most recent observational one, $\Om_M = 0.302$, of the {\it Dark Energy Survey} (DES) consortium [57], while resting quite far also from its {\it Pantheon Compilation} counterpart, $\Om_M = 0.306$ [30].  It is evident that the exact value of $\Om_M$, as also of many other parameters of cosmological significance (see, e.g., [175]), is still a matter of debate.

\subsection{The accelerated expansion of the Universe}

Upon consideration of Eq. (18),  Eq. (17) is written in the form \be \left ( \frac{H}{H_0} \right )^2 = \left ( \frac{S_0}{S} \right )^3 \left [ \Om_M + (1 - \Om_M) \left ( \frac{S}{S_0} \right )^{3 (1 - \Gm)} \right ] \ee or, it terms of the cosmic scale factor, in the more convenient form \be \left [ \frac{d}{d t} \left ( \frac{S}{S_0} \right )^{3/2} \right ]^2 = \frac{1}{t_{EdS}^2} \left \lbrace \Om_M + (1 - \Om_M) \left [ \left ( \frac{S}{S_0} \right )^{3/2} \right ]^{2 (1 - \Gm)} \right \rbrace \: , \ee where $t_{EdS} = \frac{2}{3 H_0}$ is the age of the Universe in the Einstein-de Sitter (EdS) model. Eq. (22), can be solved in terms of hypergeometric functions, as follows \be \left ( \frac{S}{S_0} \right )^{\frac{3}{2}} \: _2F_1 \left ( \frac{1}{2 (1 - \Gm)} \: , \: \frac{1}{2} \: ; \: \frac{3 - 2 \Gm}{2 (1 - \Gm)} \: ; - \left ( \frac{1 - \Om_M}{\Om_M} \right ) \left [ \frac{S}{S_0} \right ]^{3 (1 - \Gm)} \right ) = \sqrt{\Om_M} \left ( \frac{t}{t_{EdS}} \right ) \ee (cf. [176], pp. 1005 - 1008). For $\Gm < 1$, the resulting hypergeometric series converges absolutely within the circle of (unit) radius $\left \vert \frac{S}{S_0} \right \vert \leq 1$ (cf. [177], p. 556). There are two limiting cases of Eq. (23), of particular interest: \texttt{(i)} For $\Om_M = 1$, it yields $S = S_0 \left ( \frac{t}{t_{EdS}} \right )^{2/3}$, i.e., the scale factor of the EdS model. \texttt{(ii)} For $\Gm = 0$, (i.e., in the $\Lm$CDM-like limit), Eq. (23) is written in the form \be \left ( \frac{S}{S_0} \right )^{\frac{3}{2}} \: _2F_1 \left ( \frac{1}{2} \: , \: \frac{1}{2} \: ; \: \frac{3}{2} \: ; \: - \left ( \frac{1 - \Om_M}{\Om_M} \right ) \left [ \frac{S}{S_0} \right ]^3  \right ) = \sqrt{\Om_M} \left ( \frac{t}{t_{EdS}} \right ) \: , \ee which, upon consideration of the identity \be _2F_1 \left ( \frac{1}{2} \: , \: \frac{1}{2} \: ; \: \frac{3}{2} \: ; \: - x^2  \right ) = \frac{1}{x} \sinh^{-1} (x) \ee (cf. [176], Eq. 9.121.28, p. 1007 and [177], Eq. 15.1.7, p. 556), where in our case, $x = \sqrt{\left ( \frac{1 - \Om_M}{\Om_M} \right ) \left [ \frac{S}{S_0} \right ]^3}$, results in \be S(t) = S_0 \left ( \frac{\Om_M}{1 - \Om_M} \right )^{1/3} \sinh^{2/3} \left ( \sqrt{1 - \Om_M} \frac{t}{t_{EdS}} \right ) \: . \ee For $1 - \Om_M = \Om_{\Lm}$, Eq. (26) represents the scale factor of the $\Lm$CDM model (cf. Eq. 5 of [178]), as it should. On the other hand, at the present epoch, i.e., when $t = t_0$ and $S = S_0$, Eq. (23) reads \be \frac{t_0}{t_{EdS}} = \frac{1}{\sqrt {\Om_M}} \: _2F_1 \left ( \frac{1}{2 (1 - \Gm)} \: , \: \frac{1}{2} \: ; \: 1 + \frac{1}{2 (1 - \Gm)} \: ; \: - \frac{1 - \Om_M}{\Om_M} \right ) \: . \ee With the aid of Eq. (27) we can eliminate $t_{EdS}$ from Eq. (23), to obtain the scale factor of the polytropic DM model (in units of $S_0$) as a function of cosmic time (in units of $t_0$), i.e., \be \left ( \frac{S}{S_0} \right )^{3/2} \frac{_2F_1 \left ( \frac{1}{2 (1 - \Gm)} \: , \: \frac{1}{2} \: ; \: \frac{3 - 2 \Gm}{2 (1 - \Gm)} \: ; \: - \left ( \frac{1 - \Om_M}{\Om_M} \right ) \left [ \frac{S}{S_0} \right ]^{3 (1 - \Gm)} \right )}{_2F_1 \left ( \frac{1}{2 (1 - \Gm)} \: , \: \frac{1}{2} \: ; \: \frac{3 - 2 \Gm}{2 (1 - \Gm)} \: ; \: - \frac{1 - \Om_M}{\Om_M} \right )} = \frac{t}{t_0} \: . \ee The evolution of $S(t)$ (in units of $S_0$) parametrized by $\Gm < 1$, is given in Fig. 1. We observe that, in all cases, there is a value of $t < t_0$ (somewhere around $t \simeq 0.75 \: t_0$), above which, the function $S(t)$ becomes concave, i.e., $\ddot{S} > 0$. This is a very important result, indicating that the polytropic DM model with $\Gm < 1$ definitely transits from deceleration to acceleration at a certain time, (quite) close to the present epoch, $t_0$. 

\begin{figure}[ht!]
\centerline{\mbox {\epsfxsize=14.cm \epsfysize=9.cm
\epsfbox{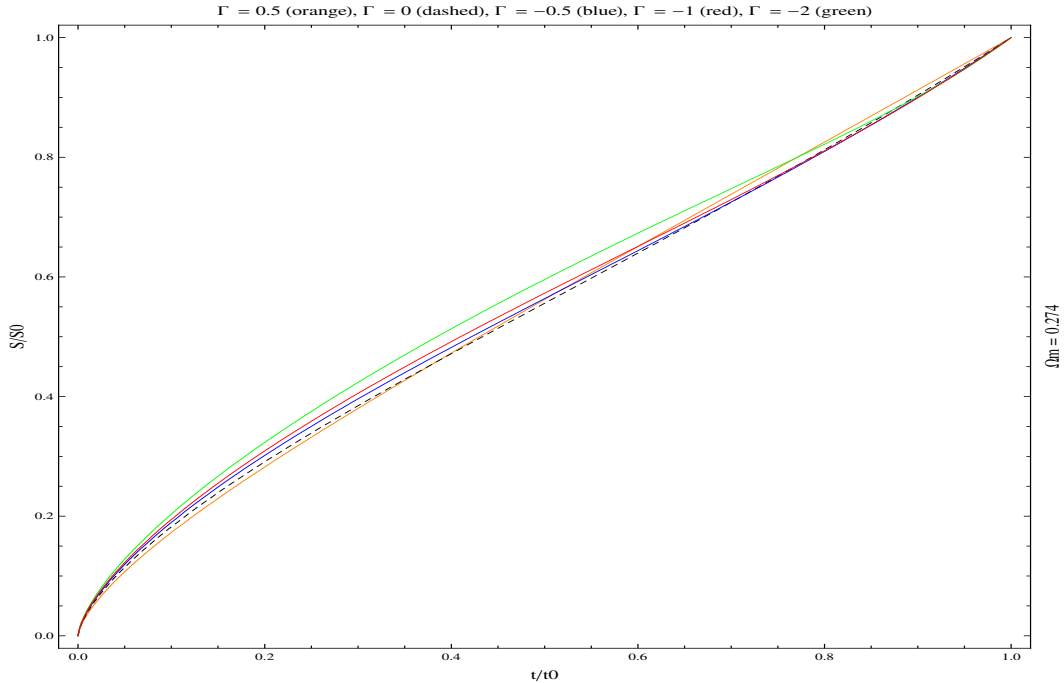}}} \caption{The scale factor, $S$, of the polytropic DM model in units of its present-time value, $S_0$, as a function of cosmic time $t$ (in units of $t_0$), for $\Gm = 0.5$ (orange), $\Gm = 0$ (dashed), $\Gm = -0.5$ (blue), $\Gm = - 1$ (red), and $\Gm = - 2$ (green). For each and every curve, there is a value of $t < t_0$ above which $S(t)$ becomes concave, i.e., the polytropic DM Universe accelerates its expansion.}
\end{figure}

\subsection{The age of the Universe}

By construction, Eq. (27), represents the age, $t_0$, of the polytropic DM Universe in units of $t_{EdS}$. The behaviour of $t_0$ as a function of the polytropic exponent $\Gm < 1$, is presented in Fig. 2. In the $\Lm$CDM-like $(\Gm = 0)$ limit, Eq. (27) yields \be t_0 = t_{EdS} \frac{1}{\sqrt{1 - \Om_M}} \sinh^{-1} \sqrt{\frac{1 - \Om_M}{\Om_M}} \: . \ee For $\Om_M = 0.274$, Eq. (29) results in $t_0 = 1.483 \; t_{EdS}$, which, adopting that $H_0 \simeq 67.5$ km/sec/Mpc (see, e.g., [54], [57]), yields $t_0 = 13.79 \: Gys$. This, theoretically predicted value of $t_0$, is in an excellent agreement with the corresponding observational result [54 - 57] for the age of the $\Lm$CDM Universe. In fact, from Fig. 2 we see that, for every $\Gm < 1$, the age of the polytropic DM model is always larger than that of its EdS counterpart, in other words, the polytropic DM model so assumed no longer suffers from what is referred to as the age problem. 

\begin{figure}[ht!]
\centerline{\mbox {\epsfxsize=14.cm \epsfysize=9.cm
\epsfbox{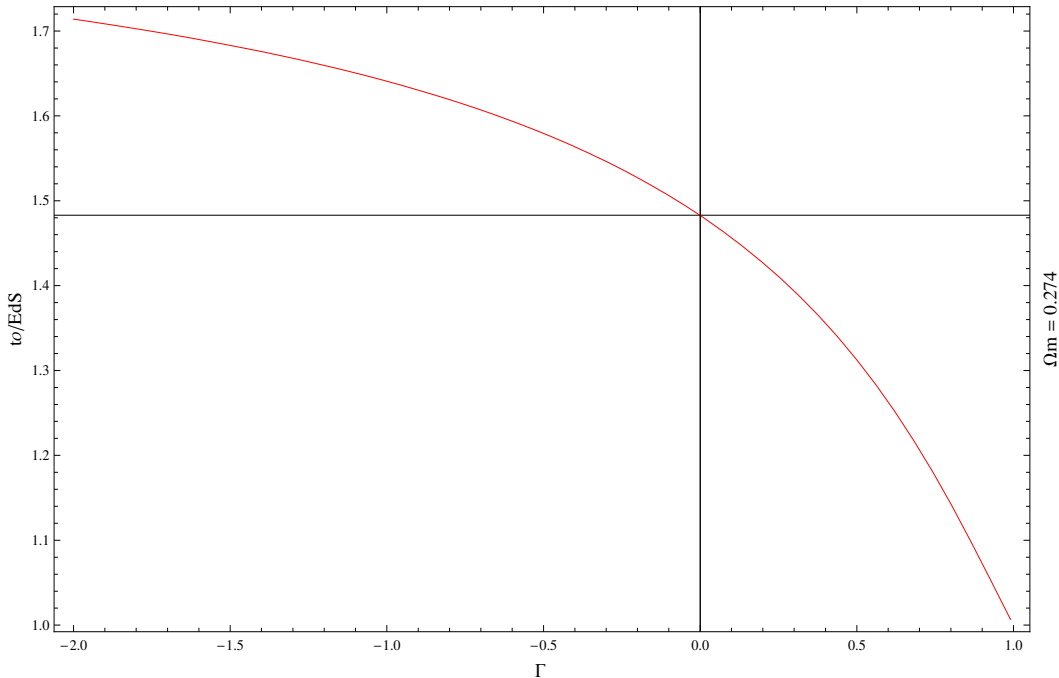}}} \caption{The age of the polytropic DM model, $t_0$, in units of $t_{EdS}$, as a function of the polytropic exponent $\Gm < 1$ (red solid line). Notice that, for every $\Gm < 1$, we have $t_0 > t_{EdS}$, with $t_0$ approaching $t_{EdS}$ only in the isothermal $(\Gm \rarrow 1)$ limit. The horizontal solid line denotes the age of the Universe in the $\Lm$CDM-like $(\Gm = 0)$ limit of the polytropic DM model, i.e., $t_0 = 1.483 \: t_{EdS}$.}
\end{figure}

\subsection{Transition to acceleration}

In the polytropic DM model under consideration, the Hubble parameter (21) in terms of the cosmological redshift, $1 + z = \frac{S_0}{S}$, is written in the form \be H = H_0 ( 1 + z )^{\frac{3}{2}} \left [ \Om_M + \frac{1 - \Om_M}{( 1 + z )^{3 (1 - \Gm)}} \right ]^{1/2} . \ee In view of Eq. (30), the deceleration parameter \be q(z) = \frac{dH/dz}{H(z)} (1+z) - 1 \ee reads \be q (z) = \frac{1}{2} \left [ 1 - \frac{3 (1 - \Gm) (1 - \Om_M)}{\Om_M (1 + z)^{3 (1 - \Gm)} + (1 - \Om_M)} \right ] \: . \ee For $z = 0$ (i.e., at the present epoch), we obtain \be q_0 = \frac{1}{2} \left [ 1 - 3 (1 - \Gm) (1 - \Om_M) \right ] \: , \ee which, in the $\Lm$CDM-like (i.e., $\Gm = 0)$ limit, yields $q_0 = - 0.54$. This result lies well within the associated observationally determined range of $q_0$, i.e., $q_0 = - 0.53_{-0.13}^{+0.15}$ [179], and, in fact, reproduces the corresponding (i.e., theoretically-derived) $\Lm$CDM result, that is, $q_0 = - 0.55 \pm 0.01$ [180]. But what is more important, is that the condition $q(z) \leq 0$ reveals a particular value of $z$, the so-called transition redshift, \be z_{tr} = \left [ (2 - 3 \Gm) \frac{1 - \Om_M}{\Om_M} \right ]^{\frac{1}{3 (1 - \Gm)}} - 1 \: , \ee below which, $q(z)$ becomes negative, i.e., the Universe accelerates its expansion. In the $\Lm$CDM-like $(\Gm = 0)$ limit, Eq. (34) yields $z_{tr} = 0.744$, which \texttt{(i)} lies well-within range of the corresponding $\Lm$CDM result, namely, $z_{tr} = 0.752 \pm 0.041$ [29] and \texttt{(ii)} actually reproduces the associated result of Muccino et al. [181], i.e., $z_{tr} = 0.739_{-0.089}^{+0.065}$, obtained by applying a model-independent method to a number of SNeIa, BAOs, and GRB data. Furthermore, by virtue of Eq. (34), the condition $z_{tr} \geq 0$ imposes a more stringent constraint on the potential values of $\Gm$, namely, \be \Gm \leq \frac{1}{3} \left [ 2 - \frac{\Om_M}{1 - \Om_M} \right ] \: . \ee For $\Om_M = 0.274$, Eq. (35) yields $\Gm \leq 0.541$. Apparently, the polytropic DM model with $\Gm \leq 0.541$ accelerates its expansion at cosmological redshifts lower than a transition value, without the need of any novel DE component. The behaviour of $z_{tr}$, as a function of the parameter $\Gm \leq 0.541$, is presented in Fig. 3. 

\begin{figure}[ht!]
\centerline{\mbox {\epsfxsize=14.cm \epsfysize=9.cm
\epsfbox{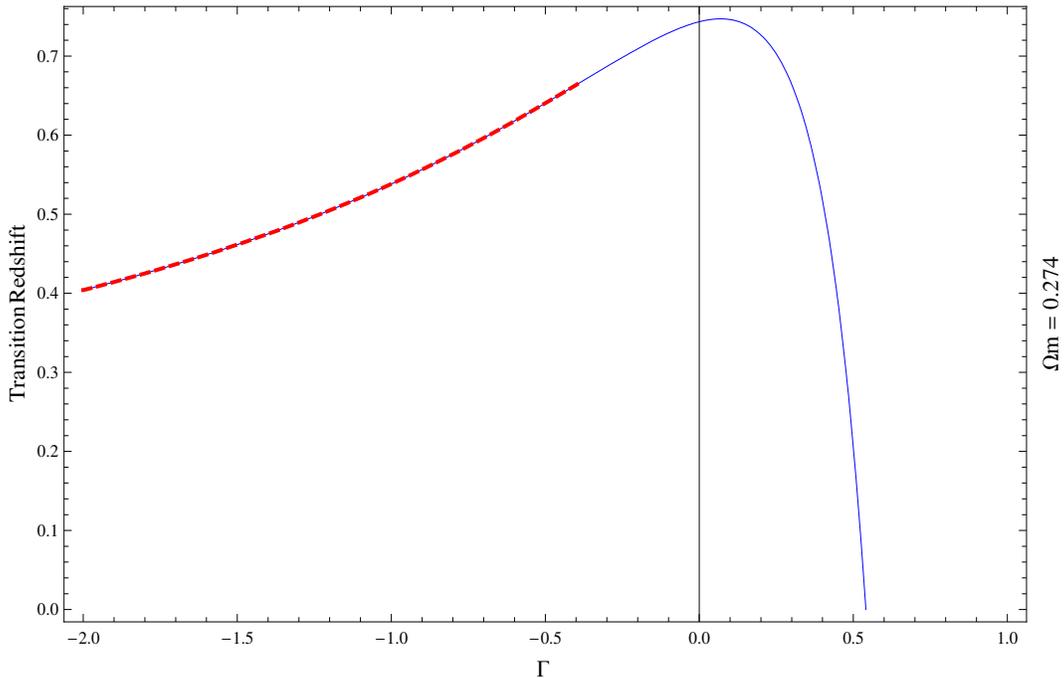}}} \caption{The transition redshift, $z_{tr}$, in the polytropic DM modelin terms of the associated exponent, $\Gm$ (blue solid curve). For $\Gm \leq - 0.38$ (red dashed curve), the Universe enters into the phantom realm [160].}
\end{figure}

\subsection{The total EoS parameter}

In the $\Lm$CDM-like $(\Gm = 0)$ limit, our model actually reproduces the behaviour of the (so-called) total EoS parameter, \be w_{tot} \equiv \frac{p}{\varep} \: , \ee as a function of $z$ [88]. For $\Gm = 0$, upon consideration of Eqs. (14), (16), and (18), Eq. (36) yields \be w_{tot} \equiv \frac{p}{\varep} = - \frac{1 - \Om_M}{1 - \Om_M + \Om_M (1 + z)^3} \: , \ee the behaviour of which, in terms of the cosmological redshift, is depicted in Fig. 4. Today, i.e., for $z = 0$, we have $w_{tot} = - \left ( 1 - \Om_M \right ) = - \Om_{\Lm}$, in complete correspondence to the $\Lm$CDM result, \be w_{tot} = \frac{p_{tot}}{\rh_{tot}} = \frac{p_{\Lm}}{\rh_M + \rh_{\Lm}} = \frac{- \rh_{\Lm}}{\rh_M + \rh_{\Lm}} = \frac{- \Om_{\Lm}}{\Om_M + \Om_{\Lm}} = - \Om_{\Lm} \ee (in connection, see, e.g., [88]).  

\begin{figure}[ht!]
\centerline{\mbox {\epsfxsize=14.cm \epsfysize=9.cm
\epsfbox{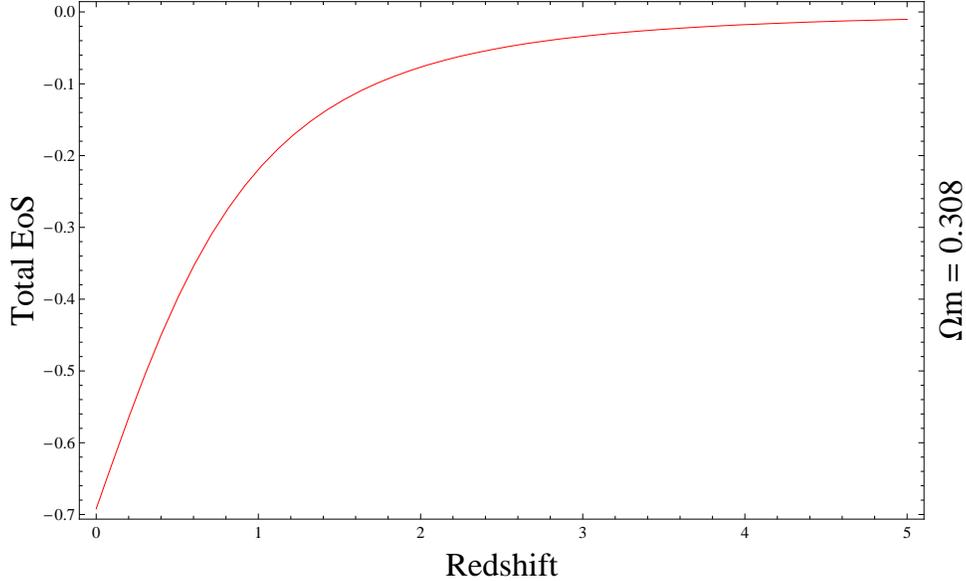}}} \caption{The total EoS parameter, $w_{tot}$,in terms of $z$, in the context of the $\Lm$CDM-like (i.e., $\Gm = 0$) limit of the polytropic DM model. Notice that, today (i.e., at $z = 0)$, $w_{tot} \approx - 0.7$, while, for larger values of $z$, it approaches zero, in complete agreement to $\Lm$CDM cosmology [88].}
\end{figure} 

\subsection{The range of values of the polytropic exponent}

The isentropic velocity of sound is defined as \be c_s^2 = c^2 \left ( \frac{\partial p}{\partial \varep} \right )_{\cal S} \ee (see, e.g., [182], p. 52), where $\left ( \frac{\partial p}{\partial \varep} \right )_{\cal S} \leq 1$, in order to avoid violation of causality [183]. In the polytropic DM model, the total energy density of the Universe matter-energy content is related to pressure by Eq. (16), whose partial differentiation yields the associated velocity of sound as a function of $z$, \be \left ( \frac{c_s}{c} \right )^2 =  - \frac{\Gm (1 - \Gm) \frac{1 - \Om_M}{\Om_M}}{(1 + z)^{3(1 - \Gm)} + \Gm \frac{1 - \Om_M}{\Om_M}} \: . \ee Now, the condition for a positive (or zero) velocity-of-sound square imposes a major constraint on $\Gm$, i.e., \be \left ( \frac{c_s}{c} \right )^2 \geq 0 \Leftrightarrow \Gm \leq 0 \: , \ee while, admitting that, today, DM is {\em cold}, i.e., at $z = 0$, \be \left ( \frac{c_s}{c} \right )^2 < \frac{1}{3} \: , \ee we obtain \be \Gm > - \frac{2}{3} \left [ \sqrt{1 + \frac{3}{4} \frac{\Om_M}{1 - \Om_M}} - 1 \right ] = - 0.1 \: . \ee Eqs. (41) and (43) significantly narrow the potential range of values of the polytropic exponent, which, from now on, rests in \be - 0.1 < \Gm \leq 0 \: . \ee Hence, in the polytropic DM model under consideration, the associated polytropic exponent, if not zero, is definitely negative and very close to zero. Notice that, in view of Eq. (44), Eq. (9) is in excellent agreement with the associated result for a generalized Chaplygin gas, $p \sim - \rh^{\al}$, arising from the combination of X-ray and SNe Ia measurements with data from Fanaroff-Riley type IIb radio-galaxies, namely, $\al = - 0.09_{-0.33}^{+0.54}$ [184]. 

\subsection{The jerk parameter}

A dimensionless third (time-)derivative of the scale factor, $S(t)$, the so-called {\em jerk parameter}, \be j (S) = \frac{1}{SH^3} \frac{d^3 S}{d t^3} \ee (see, e.g., [185, 186]), can be used to demonstrate the departure of the polytropic DM model under consideration from its $\Lm$CDM counterpart. The reason is that, for the $\Lm$CDM model $j = 1$ for every $z$. Hence, any deviation of $j$ from unity enables us to constrain the departure of the model so assumed from the $\Lm$CDM model in an effective manner [186].

In terms of the deceleration parameter, $j$ is written in the form \be j (q) = q (2 q + 1) + (1 + z) \frac{d q}{d z} \ee (see, e.g., [187]), which, in the polytropic DM model, i.e., upon consideration of Eq. (32), yields \be j (z) = 1 - \frac{9}{2} \Gm \frac{(1 - \Gm)}{1 + \frac{\Om_M}{1 - \Om_M} (1 + z)^{3 (1 - \Gm)}} \: . \ee Notice that, for $\Gm = 0$, $j = 1$; hence, once again, the $\Gm = 0$ limit of the polytropic DM model under consideration does reproduce the $\Lm$CDM model. Now, by virtue of Eq. (41), the jerk parameter (47) reads \be  j (z) = 1 + \frac{9}{2} \vert \Gm \vert \frac{(1 + \vert \Gm \vert )}{1 + \frac{\Om_M}{1 - \Om_M} (1 + z)^{3 (1 + \vert \Gm \vert )}} \: , \ee i.e., it is always positive. This is a very important result, since it guarantees that, at $z_{tr}$, a (phase) transition of the Universe expansion from deceleration to acceleration actually takes place (in connection, see [186], [188]). 

Two values of $j (z)$ are of particular interest: \texttt{(i)} Its present-time $(z = 0)$ value, given by \be j_0 \equiv j (z = 0) = 1 + \frac{9}{2} \vert \Gm \vert (1 + \vert \Gm \vert ) \: , \ee which, in view of Eq. (44) results in \be 1 \leq j_0 < 1.495 \: , \ee clearly discriminating the $\Gm \neq 0$ polytropic DM model from its $\Lm$CDM counterpart, and \texttt{(ii)} the value of the jerk parameter at transition $(z = z_{tr})$, which, upon consideration of Eq. (34), it is given by \be j_{tr} \equiv j (z_{tr}) = 1 + \frac{3}{2} \vert \Gm \vert \: . \ee In this case, we address (once again) to Muccino et al. [181] to use the corresponding model-independent constraints on $j_{tr}$, in order to estimate the value of the polytropic index, $\vert \Gm \vert$, in a model-independent way. Accordingly, adopting the best-fit value $j_{tr} = 1.028$ of [181], obtained by means of the DHE method (see [188]), Eq. (51) yields $$ \vert \Gm \vert = 0.02 \: , $$ while, adopting the corresponding DDPE value [188], $j_{tr} = 1.041$, Eq. (51) results in $$ \vert \Gm \vert = 0.03 \: . $$ Both values, not only favour a $\Gm \neq 0$ polytropic DM model, but also, are well-within range of Eq. (44), i.e., once again, compatibility of the polytropic DM model with observation is well established.

In view of [186], we cannot help but wondering whether the polytropic DM model with a jerk parameter given by Eq. (48) is also compatible to the Union 2.1 Compilation of the SNeIa data or not.

\subsection{The Hubble diagram of the SNe Ia data}

Today, (too) many samples of SNe Ia data are used to scrutinize the viability of the DE models proposed. One of the most extended is the Union 2.1 Compilation [29], consisting of 580 SNe Ia events, being inferior only to the (so-called) Pantheon Compilation [30]. We shall use the former sample to demonstrate compatibilty of the theoretically derived (in the context of the polytropic DM model) formula for the distance modulus, \be \mu(z) = 5 \log \left ( \frac{d_L}{Mpc} \right ) + 25 \ee (see, e.g., [173], Eqs. 13.10 and 13.12, p. 359), where \be d_L (z) = c (1+z) \int_0^z \frac{d z^{\prime}}{H(z^{\prime})} \ee is the luminosity distance of a light source measured in megaparsecs (see, e.g., [189], p. 76), with the observationally determined Hubble diagram of the SNe Ia standard candles [29]. 

Upon consideration of  Eq. (30), Eq. (53) results in (see, e.g., [176], pp. 1005 - 1008) \bea && d_L (z) = \frac{2 c}{H_0} \frac{1}{\sqrt {1 - \Om_M}} \frac{1 + z}{2 - 3 \Gm} \left [ (1 + z)^{\frac{2 - 3 \Gm}{2}} \times \right . \nn \\ && \left . _2F_1 \left ( \frac{2 - 3 \Gm}{6 ( 1 - \Gm)}  \: , \: \frac{1}{2} \: ; \: \frac{8 - 9 \Gm}{6 ( 1 - \Gm)} \: ; \: - \left [ \frac{\Om_M}{1 - \Om_M} \right ] (1 + z)^{3 (1 - \Gm)} \right )  - \right . \nn \\ && \left . _2F_1 \left ( \frac{2 - 3 \Gm}{6 (1 - \Gm)}  \: , \: \frac{1}{2} \: ; \: \frac{8 - 9 \Gm}{6 ( 1 - \Gm)} \: ; \: - \left [ \frac{\Om_M}{1 - \Om_M} \right ] \right ) \right ] \: , \eea where, once again $_2F_1$ is Gauss hypergeometric function. Using Eq. (54), we overplot $\mu (z)$ on the Hubble diagram of the Union 2.1 Compilation [29] to obtain Fig. 5. We see that, in the polytropic DM model under consideration, the various theoretical curves representing the distance modulus fit the entire Union 2.1 dataset quite accurately. In other words, there is absolutely no disagreement between the theoretical prediction of the SNe Ia distribution in the polytropic DM model so assumed and the corresponding observational result. 

\begin{figure}[ht!]
\centerline{\mbox {\epsfxsize=14.cm \epsfysize=9.cm
\epsfbox{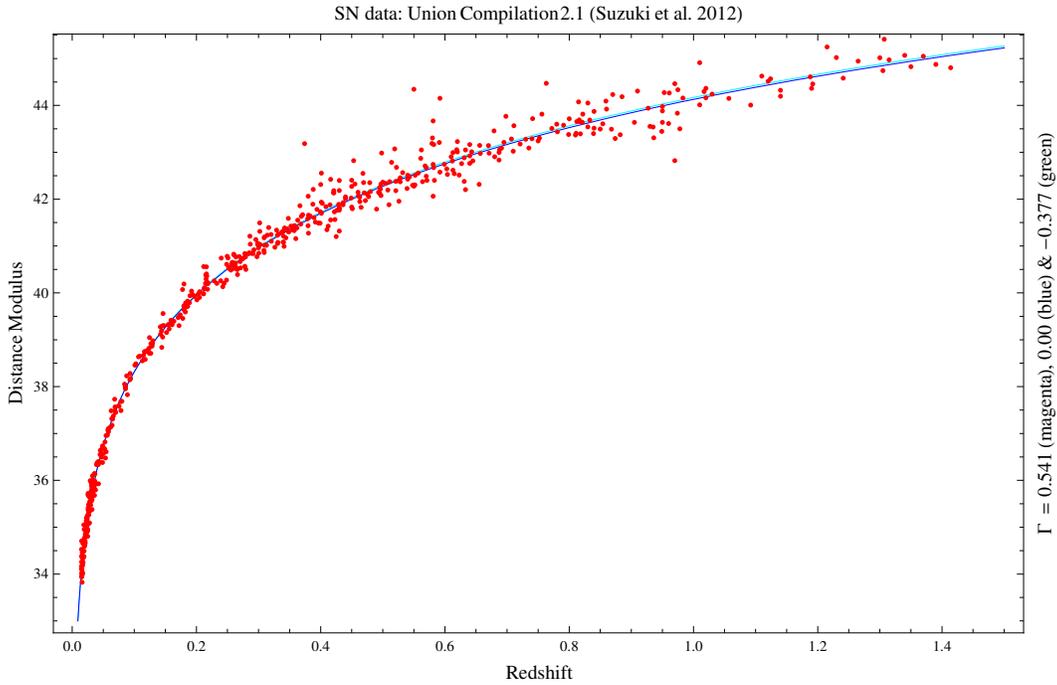}}} \caption{Hubble diagram of the Union 2.1 Compilation SNeIa data. Overplotted are three theoretically-determined curves representing the distance modulus in the polytropic DM model, i.e., Eq. (54).}
\end{figure}

\subsection{The CMB shift parameter}

The CMB shift parameter, ${\cal R}$, is widely used as a probe of DE, due to the fact that different cosmological models can result in an almost identical CMB power spectrum, if they have identical values of ${\cal R}$ [190]. For a spatially-flat cosmological model, the CMB shift parameter is given by \be {\cal R} = \sqrt{\Om_M} \int_0^{z_{*}} \frac{dz}{H(z)/H_0} \: , \ee where $z_{*}$ is the value of the cosmological redshift at photon decoupling. In the polytropic DM model under consideration, i.e., by virtue of Eq. (30), Eq. (55) is written in the form \be {\cal R} = \int_0^{z_{*}} \frac{\left ( 1 + z^{\prime} \right )^{\frac{3}{2} \vert \Gm \vert} d z^{\prime}}{\left [ (1 - \Om_M) + \Om_M \left ( 1 + z^{\prime} \right )^{3 (1 + \vert \Gm \vert )} \right ]^{1/2}} \: , \ee which, in terms of hypergeometric functions (see e.g., [176], pp. 1005 - 1008), reults in \bea && {\cal R} = \frac{2}{\left ( 2 + 3 \vert \Gm \vert \right ) \sqrt {1 - \Om_M}} \left [ (1 + z_{*})^{\frac{2 + 3 \vert \Gm \vert}{2}} \times \right . \nn \\ && \left . _2F_1 \left ( \frac{2 + 3 \vert \Gm \vert}{6 ( 1 + \vert \Gm \vert )}  \: , \: \frac{1}{2} \: ; \: \frac{8 + 9 \vert \Gm \vert}{6 ( 1 + \vert \Gm \vert )} \: ; \: - \left [ \frac{\Om_M}{1 - \Om_M} \right ] (1 + z_{*})^{3 (1 + \vert \Gm \vert )} \right . \right ) - \nn \\ && \left . _2F_1 \left ( \frac{2 + 3 \vert \Gm \vert }{6 (1 + \vert \Gm \vert )}  \: , \: \frac{1}{2} \: ; \: \frac{8 + 9 \vert \Gm \vert }{6 ( 1 + \vert \Gm \vert )} \: ; \: - \left [ \frac{\Om_M}{1 - \Om_M} \right ] \right ) \right ] . \eea To determine the value of ${\cal R}$, we adopt the {\it nine-year WMAP survey} result [191], that, $z_{*} = 1091.64 \pm 0.47$. Accordingly, for $\Gm = 0$, Eq. (57) yields \be {\cal R} = 1.7342 \: , \ee while, according to the {\it nine-year WMAP survey} [191], the value of the shift parameter in the standard $\Lm$CDM cosmology is given by \be {\cal R} = 1.7329 \pm 0.0058 \: . \ee In other words, the theoretical value of the shift parameter in the $\Lm$CDM-like limit of the polytropic DM model reproduces, to high accuracy, the corresponding result obtained by fitting the CMB data to the standard $\Lm$CDM model; hence, in the limit of $\Gm = 0$, the polytropic DM model under consideration may very well reproduce also the observed CMB spectrum. 

\subsection{The spectral index of cosmological perturbations}

The dimensionless power spectrum of rest-mass density perturbations in an isotropic Universe is defined as \be \Dl^2(\dl) = \frac{1}{2 \pi^2} k^3 \vert \dl (k) \vert^2 , \ee where $\dl = \frac{\dl \rh}{\rh}$ is the density contrast and $k$ is the associated wavenumber (see, e.g., [189], pp. 464-469). In a similar fashion, the metric counterpart of Eq. (60) is given by \be \Dl^2(\phi) = \frac{1}{2 \pi^2} k^3 \vert \phi(k) \vert^2 , \ee where $\phi$ denotes the perturbation around a spatially-flat metric [162]. Usually, $\Dl^2 (\dl)$ is parametrized as \be \Dl^2 (\dl) \sim k^{3 + n_s} \ee (see, e.g., [192], pp. 291, 292), where $n_s$ is the scalar spectral index [193]. Once again, we can test the validity of the polytropic DM model by reproducing the spectrum of rest-mass density perturbations in the associated $\Lm$CDM-like limit. The reason is that, most of the observational data accumulated so far, are model dependent [175] and, currently, the most popular model is the so-called concordance, i.e., $\Lm$CDM model [13]. Accordingly, as regards the dimensionless power spectrum of cosmological perturbations in the $\Lm$CDM-like limit of the polytropic DM model under consideration, we have \be \frac{\Dl^2 (\dl)}{\Dl^2 (\phi)} = 4 \left [ 1 + \frac{1}{3} \left ( \frac{k_{ph}}{H} \right )^2 \right ]^2 \: , \ee where $k_{ph} = k/S(t)$ is the associated physical wavenumber [162]. The behaviour of Eq. (63) as a function of $k_{ph}$ (in units of $H$) is depicted in Fig. 6 (red solid line). Accordingly, we observe that for $\left ( \frac{k_{ph}}{H} \right ) \geq 5$, i.e., for every physical wavelength less than the horizon length (dashed verical line), the quantity $\Dl^2 (\dl) / \Dl^2 (\phi)$ exhibits a prominent power-law dependence on $k_{ph}$, of the form \be \frac{\Dl^2 (\dl)}{\Dl^2 (\phi)} \sim \left ( \frac{k_{ph} }{H} \right )^{3.970} \ee and, therefore, \be \Dl^2 (\phi) \sim \frac{\Dl^2 (\dl)}{\left ( \frac{k_{ph}}{H} \right )^{3.970}} = \frac{\left ( \frac{k_{ph}}{H} \right )^{n_s + 3}}{\left ( \frac{k_{ph}}{H} \right )^{3.970}} = \left ( \frac{k_{ph}}{H} \right )^{n_s - 0.970} . \ee CMB anisotropy measurements (see, e.g., [52, 53]) and several physical arguments (see, e.g., [189], p. 466, [192], p. 292) suggest that the power spectrum of metric perturbations is scale invariant, i.e., $\Dl^2 (\phi) \sim k^0$. In this case, Eq. (65) yields \be n_s = 0.970 \: . \ee In view of Eqs. (62) and (66), we see that, although in principle there is no reason why the rest-mass density spectrum should exhibit a power-law behaviour, in the context of the polytropic DM model it effectively does so, i.e., \be \Dl^2 (\dl) \sim k_{ph}^{3 + n_s^{eff}}, ~\mbox{with} ~~ n_s^{eff} = 0.970 \: . \ee What is more important, is that, the theoretically derived value (67) for the effective scalar spectral index of rest-mass density perturbations in the $\Lm$CDM-like limit of the polytropic DM model, actually reproduces the corresponding observational (i.e., {\em Planck}) result, $n_s^{obs} = 0.968 \pm 0.006$ [55, 56]. In short, matter perturbations of linear dimensions smaller than the Hubble radius, when considered in the $\Lm$CDM-like (i.e., $\Gm = 0$) limit of the polytropic DM model under consideration, effectively exhibit a power-law behaviour of the form $\vert \dl \vert^2 \sim k^{n_s^{eff}}$, with the associated scalar spectral index being equal to $n_s^{eff} = 0.970$, i.e., very close to observation.

\begin{figure}[ht!]
\centerline{\mbox {\epsfxsize=14.cm \epsfysize=9.cm
\epsfbox{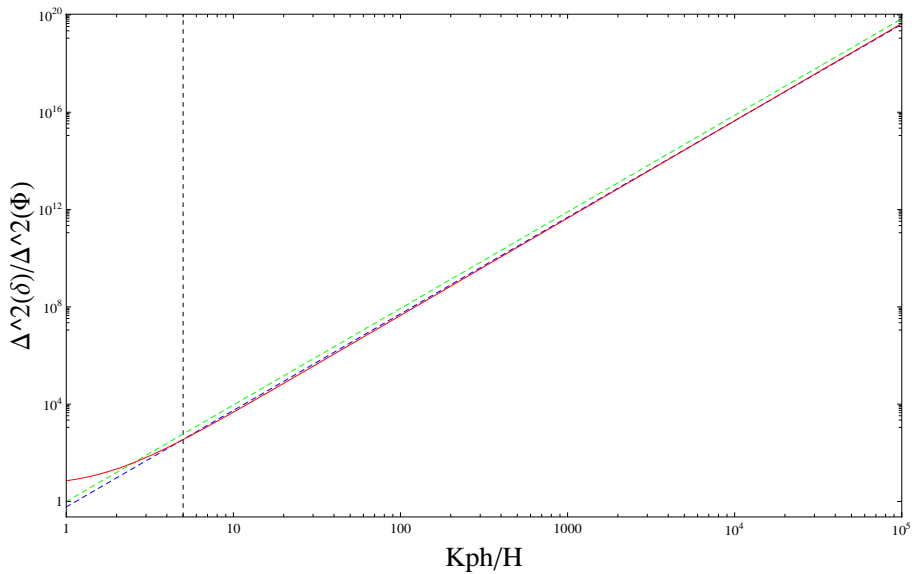}}} \caption{Small-scale perturbations i.e., Eq. (63), in the $\Gm = 0$ limit of a polytropic DM model (red solid line). The straight dashed lines, each one of slope $\al = 3.970$ represent Eq. (64); hence, in this model, rest-mass density perturbations with $\left ( \frac{k_{ph}}{H} \right ) \geq 5$ exhibit an effective power-law behaviour with a scalar spectral index equal to $n_s^{eff} = 0.970$.}
\end{figure}

\subsection{Rest-mass energy - DE equality}

In view of Eq. (19), the rest-mass energy density, $\varep_{mat} = \rh c^2$, and the internal (dark) energy density, $\varep_{int} = \varep - \varep_{mat}$, of the polytropic DM model under consideration satisfy the relation \be \frac{\varep_{int}}{\varep_{mat}} = \frac{1 - \Om_M}{\Om_M} \frac{1}{(1+z)^{3(1 - \Gm)}} \: . \ee Eq. (68) suggests that, for $\Gm = 0$, DE becomes equal to its rest-mass counterpart, not at transition $(z_{tr} = 0.744)$, but, quite later, at $z_{eq} = 0.384$, which is very close to the corresponding observationally determined value $z_{eq} = 0.391 \pm 0.033$ [29], associated (once again) to $\Lm$CDM model. 

\subsection{It is not a coincidence}

The evolution of a spatially-flat FRW model is governed by Eqs. (5), (6), and (8). The combination of them results in \be \frac{\ddot{S}}{S} = - \frac{4 \pi G}{3 c^2} \left ( \eps + 3 p \right ) \ee (see, e.g., [43, 44]); hence, the condition for accelerated expansion, $\ddot{S} > 0$, yields \be \varep + 3 p < 0 \: . \ee In the context of the polytropic DM model, condition (70) is written in the form \be \rh_0 c^2 (1 + z)^3 \left [ 1 - (2 + 3 \vert \Gm \vert) \frac{1 - \Om_M}{\Om_M} \frac{1}{(1 + z)^{3(1+ \vert \Gm \vert)}} \right ] < 0 \: , \ee in view of which, such a model accelerates its expansion at cosmological redshifts lower than a particular value, namely, \be z < \left [ (2 + 3 \vert \Gm \vert) \frac{1 - \Om_M}{\Om_M} \right ]^{\frac{1}{3(1+ \vert \Gm \vert)}} - 1 \equiv z_{tr} \: , \ee in complete correspondence to Eq. (34). According to Eqs. (70) and (72), the assumption that the cosmological evolution can be driven by a polytropic DM fluid could most definitely explain why the Universe transits from deceleration to acceleration at $z_{tr}$, without the need for any novel DE component or the cosmological constant. Instead, it would reveal a conventional form of DE, i.e., the one due to this fluid's internal motions, which, so far, has been disregarded [113].

\section{Discussion and Conclusions}

The possibility that the extra DE needed to compromise both spatial flatness and the accelerated expansion of the Universe actually corresponds to the thermodynamic internal energy of the cosmic fluid itself is reviewed and further scrutinized. In this approach, the Universe is filled with a perfect fluid of collisional DM, the volume elements of which perform polytropic flows [160 - 163]. In the distant past $(z \gg 1)$ the polytropic DM model so assumed behaves as an EdS model, filled with dust (cf. Eq. 32), while, on the approach to the present epoch $(t \simeq 0.75 \: t_0)$, the internal physical characteristics of the cosmic fluid take over its dynamics (cf. Eq. 68). Their energy can compensate the DE needed to compromise spatial flatness (cf. Eq. 20), while, the associated cosmic pressure is negative (cf. Eq. 18). As a consequence, the polytropic DM model under consideration accelerates its expansion at cosmological redshifts lower than a transition value (cf. Eq. 34), in consistency with condition $\varep + 3 p < 0$ (cf. Eq. 72). This model is characterized by a free parameter, the associated polytropic exponent $\Gm$. In fact, several physical arguments can impose successive constraints on $\Gm$, which, eventually, settles down to the range $- 0.1 < \Gm \leq 0$ (cf. Eq. 44), namely, if it is not zero (i.e., a $\Lm$CDM-like model), it is definitely negative and very close to zero. 

 The polytropic DM model under consideration can reproduce all the major observational results of conventional (i.e., $\Lm$CDM) Cosmology, simply by means of a single fluid, i.e., without {\em a priori} assuming the existence of any DE component and/or the cosmological constant. This model actually belongs to the broad class of the {\em unified DE models}, in which the DE effects are due to the particular properties of the (unique) cosmic fluid (in connection, see, e.g., [194, 195]). 

We can test the validity of the polytropic DM model so assumed, by reproducing all the current cosmological issues in the associated $\Lm$CDM-like limit. The reason is that, most of the observational data accumulated so far are model dependent [175] and, currently, the most popular model is the $\Lm$CDM model. In this context, our polytropic DM model can confront all major issues of cosmological significance, as, e.g., 

\begin{itemize}

\item The nature of the universal (dark) energy deficit needed to compromise spatial flatness: In the polytropic DM model under consideration it can be attributed to thermodynamic energy of the associated fluid internal motions (cf. Eqs. 19 and 20).

\item The accelerated expansion of the Universe: For $t > 0.75 \; t_0$ (i.e., quite close to the present epoch), the solution of Friedmann equation that governs the evolution of the scale factor, $S(t)$,  in the polytropic DM model, becomes concave, i.e., $\ddot{S} > 0$ resulting in the acceleration of the Universe expansion (see, e.g., Fig. 1).

\item The age problem: For every $- 0.1 < \Gm \leq 0$, the age of the polytropic DM model, $t_0$, is always larger than that of its EdS counterpart, $t_{EdS} = \frac{2}{3 H_0}$. In the $\Lm$CDM-like $(\Gm = 0)$ limit, we obtain $t_0 = 1.483 \: t_{EdS} = 13.79 \: Gys$, in complete agreement with the corresponding observational result [55 - 57] for the age of the $\Lm$CDM Universe (see, e.g., Fig. 2).

\item The value of the cosmological redshift parameter at which transition from deceleration to acceleration takes place, $z_{tr}$: In the $\Lm$CDM-like limit (i.e., $\Gm = 0$) of the polytropic DM model so assumed, we obtain $z_{tr} = 0.744$ (cf. Fig. 3), which lies well-within range of the corresponding $\Lm$CDM result, namely, $z_{tr} = 0.752 \pm 0.041$ [29],  as well as in the associated model-independent range $z_{tr} = 0.739_{-0.089}^{+0.065}$ [181].

\item The long-sought theoretical value of the deceleration parameter, $q$, at the present epoch: In the $\Lm$CDM-like limit of the polytropic DM model under consideration, $q_0 = - 0.54$ (cf. Eq. 33, for $\Gm = 0$), that is fully compatible with the observational result, $q_0 = - 0.53_{-0.13}^{+0.17}$ [179], associated to the $\Lm$CDM model.  

\item The behaviour of the total EoS parameter, $w$: In the $\Lm$CDM-like (i.e., $\Gm = 0$) limit of the polytropic DM model, today, $w_{tot} \approx - 0.7$ (cf. Fig. 4), while, as $z$ grows, $w_{tot} \rarrow 0$, as suggested by $\Lm$CDM cosmology [88].

\item The resulting range of values of the polytropic index, $ - 0.1 < \Gm \leq 0$: It is in excellent agreement with the associated result for a generalized Chaplygin gas, $p \sim - \rh^{\al}$, arising from the combination of X-ray and SNe Ia measurements with data from Fanaroff-Riley type IIb radio-galaxies, namely, $\al = - 0.09_{-0.33}^{+0.54}$ [184]. 

\item  The behaviour of the associated jerk parameter, $j(z)$: The polytropic DM model possesses a positive jerk parameter, with the aid of which (at transition) we can also estimate the value of the polytropic index, $\vert \Gm \vert$, in a model-independent manner [181], namely, $\vert \Gm \vert \in (0.02, \: 0.03)$.  

\item The Hubble diagram of the SNe Ia standard candles: In the polytropic DM model under consideration, the theoretically derived distance modulus fits the entire Union 2.1 dataset [29] with accuracy. In other words, there is absolutely no disagreement between the theoretical prediction of our model and the observed distribution of the distant SNe Ia events (cf. Fig. 5).

\item The CMB shift parameter: In the $\Lm$CDM-like limit of the polytropic DM model, ${\cal R} = 1.7342$, while, according to the {\it nine-year WMAP survey}, the value of the CMB shift parameter in the standard $\Lm$CDM model is $ {\cal R} = 1.7329 \pm 0.0058$ [191]. In other words, the value of the CMB shift parameter in the $\Lm$CDM-like limit of the polytropic DM model actually reproduces the corresponding result obtained by fitting the CMB data to the standard $\Lm$CDM model. It is, therefore, expected that, in the limit $\Gm = 0$, the polytropic DM model under consideration may very well reproduce also the observed CMB spectrum. 

\item And, in fact, it actually does so (cf. Eq. 67), since the theoretically derived value for the effective scalar spectral index of rest-mass density perturbations in the $\Lm$CDM-like limit of the polytropic DM model, $n_s^{eff} = 0.970$, actually reproduces the corresponding observational {\it Planck} result, $n_s^{obs} = 0.968 \pm 0.006$ [55, 56]. In other words, matter perturbations of linear dimensions smaller than the horizon length, when considered in the $\Lm$CDM-like (i.e., $\Gm = 0$) limit of a polytropic DM model, effectively exhibit a power-law behaviour of the form $\vert \dl \vert^2 \sim k^{n_s^{eff}}$, with the associated scalar spectral index being equal to $n_s^{eff} = 0.970$, i.e., very close to observation.

\item The rest-mass energy - DE equality: In the $\Lm$CDM-like limit of the polytropic DM model under consideration (cf. Eq. 68), DE becomes equal to its rest-mass counterpart at $z_{eq} = 0.384$, which is remarkably close to the corresponding observationally determined value $z_{eq} = 0.391 \pm 0.033$ [29], associated to $\Lm$CDM model.

\item Finally, the polytropic DM model can, most definitely, explain why the Universe transits to acceleration at $z_{tr}$, without the need for any novel DE component or the cosmological constant, solely being consistent with the general relativistic condition, that, $\varep + 3 p < 0$ (cf. Eqs. 70 and 72).

\end{itemize}

Compatibility of the polytropic DM model with the observational constraints on all the parameters of cosmological significance needs to be further explored and scrutinized, in order to decide on the likelihood of this model over all other alternatives and, especially, the $\Lm$CDM model. Clearly, the ultimate verification of any (unified or not) DE model would be the reproduction of the observed DM halo distributions and the associated galactic evolution. In this context, preliminary results regarding the evolution of small-scale density perturbations at low redshift values, suggest that, in the $c_s^2 \neq 0$ case of the polytropic DM model, the density-contrast profile, $\dl (z)$, consists of {\em peaks and troughs} that resemble the observed galaxy distribution (in $z$). Therefore, as regards the evolution of small-scale density perturbations in a polytropic DM model with $c_s^2 \neq 0$, a more elaborated study is necessary and it will be the scope of a future work.

Finally, it is clear that this review article neither deals with nor takes into account the fundamental nature of the polytropic DM constituents, i.e., the field nature of the cosmic fluid. In this context, recent studies suggest that certain barotropic fluids may arise naturally from a $k-$essence lagrangian, involving a self-interacting (real or complex) scalar field [196]. In direct connection to the quantum origin of the polytropic DM fluid, one should also address the origin of the (extra) amount of {\em heat}, ${\cal C} dT$, offered to the volume elements, as suggested by Eq. (7). According to [76], this could be due to a long-range confining force between the DM particles. In our case, it would be of the form $F = - K r^{2 + 3 \vert \Gm \vert}$, where $r$ is the radial distance and $K > 0$ is a normalization constant (in connection, see Eqs. (80) and (89) of [76]). This force may be either of gravitational origin or a new force [141], [144]. However, it is not yet clear whether a system subject to a long-range confining force can reach thermodynamic equilibrium, hence, this is also a matter of debate that must be addressed in future studies.

In any case, instead of treating any novel DE component and/or modified gravity theories as pillars of contemporary Cosmology, let us address a much simpler possibility: The polytropic flow of the conventional matter-energy content of the Universe, in connection to a potential self-interacting nature of DM [197]. As we have demonstrated in this review, the yet ignored thermodynamical content of the Universe could arise as a mighty and relatively inexpensive contestant for an extra (dark) energy candidate that could compensate both spatial flatness and accelerated expansion. In view of all the above, the cosmological model with matter content in the form of a self-interacting DM fluid whose volume elements perform polytropic flows looks very promising and should be further explored and scrutinized in the search for a viable alternative to $\Lm$CDM model.







%

\end{document}